\documentclass[sn-mathphys]{sn-jnl}

\jyear{2021}%

\theoremstyle{thmstyleone}%
\theoremstyle{thmstyletwo}%

\theoremstyle{thmstylethree}%

\def\eq#1{(\ref{#1})}

\begin{document}

\title[Article Title]{Spacetime from quantum information: spin networks and the cosmological constant in the $AdS/CFT$ correspondence}


\author*[1]{\fnm{Carlos} \sur{Silva}}\email{carlosalex.phys@gmail.com}

%

\affil*[1]{\orgname{Instituto Federal de Educa\c{c}\~{a}o Ci\^{e}ncia e Tecnologia do Cear\'{a} (IFCE)}, Campus Tiangu\'{a} \orgaddress{\street{Av. Tabeli\~{a}o Luiz Nogueira de Lima, s/n - Santo Ant\^{o}nio}, \city{Tiangu\'{a}}, \postcode{62320-000}, \state{Cear\'{a}}, \country{Brazil}}}

%


\abstract{Based on a recently proposed holographic relationship between spin networks and superstrings, we find out numerical values for the matter and cosmological constant density parameters given by $\Omega_{m0} = 0.3311$ and $\Omega_{\Lambda 0} = 0.6689$, respectively. Such values are in good approximation with the Planck 2018 cosmological data \cite{Aghanim:2018eyx}. Moreover, by using the current value of the Hubble parameter given by Planck 2018 collaboration, we obtain $\Lambda_{0} = 2.779 \times 10^{-122}$ as an estimation of the cosmological constant. Such results can be seen as possible phenomenological evidence for quantum gravity, relying on what may be a sui-generis feature of it: that spacetime must emerge from quantum information.}




\maketitle

\section{Introduction}\label{sec1}

One of the most intricate problems in modern physics consists of the fact that the observed value of the cosmological constant is much smaller than that given by particle physics (about 122 orders of magnitude). Such an unconformity between theoretical results and observations has been described as "the largest discrepancy between theory and experiment in all of science" \cite{Adler:1995vd}, and as the " worst theoretical
prediction in the history of physics" \cite{Hobson:2006se}. In addition, the fine balance between the cosmological constant and matter in our universe, in the present epoch, consists of another mystery. This is called the cosmological coincidence problem \cite{Weinberg:2000yb, Velten:2014nra}. 

The two troubles above, known as the "old"  and the "new" cosmological constant problem \cite{Weinberg:2000yb}, respectively, have consequences for different aspects of our universe, e.g, for the structure formation and consequently for the existence of life \cite{Weinberg:1987dv, Weinberg:2000yb, Velten:2014nra}.
In face of them, a universe like ours, with conditions to support life, would be very improbable, and a fine-tuning mechanism for its initial conditions becomes necessary \cite{Weinberg:1987dv}.
The requisite of such a fine-tuning has launched the idea that must exist not only our universe, but a multiverse with different values for the cosmological constant, and that we have had the luck of stay in a universe with a value that supports life. Such an idea has been supported, for example, by the string landscape \cite{Polchinski:2006gy, Bousso:2011izg}.

It is expected that a theory of quantum gravity must shed some light on the issue of the cosmological constant \cite{Weinberg:1988cp}. However, the road to a quantum description of the gravitational phenomena has been a tough challenge.
In this sense, the main approaches for such a purpose, superstring theory \cite{Becker:2007zj}, and loop quantum gravity (LQG) \cite{Rovelli:2007uwt},
seem to address different aspects of reality.

In LQG, the fundamental objects are spin networks describing a pre-geometric regime, in a way that LQG consists of a background independent treatment to quantum gravity, in line with the theory of general relativity. 
While LQG does not unify gravity with matter, in string theory, matter as well as the fundamental interactions can be set out in a unifying way through the vibrations of strings that move within a pre-established classical spacetime. Particularly, the gravitational force is depicted by the vibrations of closed strings.
Other conflicts arise: string theory requires that spacetime must have $10$ dimensions, and needs the existence of supersymmetry. LQG, on the other side, is a four-dimensional theory, and supersymmetry is not a feature of it.

In addition to the discrepancy between string theory and LQG, the lack of any phenomenological evidence for both of them may lead such theories to a crisis scenario.
In fact, it can be very difficult to obtain any experimental evidence for such approaches, e.g., from particle accelerators, since building such kind of device with enough energy to access the scales where string theory and LQG become relevant would be unthinkably expensive. 

In front of such difficulties, it has been interesting to turn our attention to cosmology and use the universe as a laboratory \cite{Barrau:2017tcd}. In this sense, the correct prediction of the $\Omega_{\Lambda_{0}}$, $\Omega_{m_{0}}$ and cosmological constant values, due to its relevance, could be presented as an important piece of evidence for any candidate theory to quantum gravity.
However, so far no satisfactory theoretical insight to accomplish such a task has appeared in string theory, LQG, nor in another theoretical approach. 



On the other hand, recently a new route for quantum gravity has been paved by a version of the $AdS/CFT$ correspondence where closed strings can be holographically related to spin networks \cite{Silva:2020bnn}. Such a bridge between string theory and LQG has been useful to address some known problems related to quantum gravity research. For example, in such a scenario, the stumbling block of the big bang singularity in $AdS/CFT$ \cite{Bak:2006nh, Engelhardt:2015gla} has been solved, and it has been possible to circumvent the infamous LQG Imirzi ambiguity \cite{Rovelli:1997na}.

In the present work, we shall use the results obtained in \cite{Silva:2020bnn} to undertake an interesting idea that has sprung up in the context of both string theory and LQG: that spacetime must emerge from quantum information. It will lead us to a relational approach where branes can be conceived as quantum reference frames from which spacetime must emerge.  Possible observational evidence for such a scenario can be presented: the prevision of values for the cosmological constant and matter density parameters  in  good approximation with the Planck 2018 results \cite{Aghanim:2018eyx}. Moreover, by using the current value of the Hubble parameter given by Planck 2018 collaboration, a value for the cosmological constant that is in agreement with observations has also been obtained.

The article is organized as follows: in section \eq{sec2}, we shall review the main aspects of the holographic relationship between string theory and LQG traced out in \cite{Silva:2020bnn}. In section \eq{sec3}, by implementing the idea that spacetime must emerge from quantum information, we shall generalize the results of  \cite{Silva:2020bnn} to the case of $N$ branes. In section \eq{sec4}, we shall demonstrate that the quantum gravity description of a system of $N$ branes is given by full (abstract) spin networks. In section \eq{sec5}, we shall apply such results to the issue of the cosmological constant.
Section \eq{conc} is devoted to conclusions and discussions. Throughout the paper, we have used $G = \hslash = c = 1$.

\section{- Braneworld spin networks and closed strings.}\label{sec2}

D-branes are the main objects that appear in the non-perturbative sector of string theory and have a very important role in the context of the $AdS/CFT$ correspondence \cite{Ammon:2015wua}.
Results introduced in \cite{Silva:2020bnn} can however lead such objects to gain additional status in the $AdS/CFT$ scenario,
by showing that D-branes can be used as a bridge between string theory and LQG.

In this way,  it has been demonstrated that the Hamiltonian operator, corresponding to a $4D$ flat FRLW universe living on the $AdS$ boundary, described by a Randall-Sundrum II brane, is given by \cite{Silva:2020bnn}

\begin{equation}
\widehat{\mathcal{H}}_{grav}  = \frac{-3V}{32\pi  \xi^2}[2\mathbb{I} - \hat{h}_{+} -  \hat{h}_{-}] \;\;, \label{q-lqc-hamiltonian}
\end{equation}

\noindent where 

\begin{equation}
\xi = (3/(16\pi \sigma))^{1/2}, \label{xi-sigma}
\end{equation}

\noindent with $\sigma$ giving the brane tension. In the Eq. \eq{q-lqc-hamiltonian}, $\mathbb{I}$ corresponds to the identity matrix.


The novelty related to the Hamiltonian above is that the operators  $\hat{h}_{\pm}$ consist of holonomies, which act as $U(1)$ transformations creating the braneworld quantum geometrical states $\psi_{x_{n}} = e^{ipx_{n}}$ as:

\begin{equation}
\hat{h}_{\pm}\psi_{x} = e^{\pm i\sqrt{\Delta} p}e^{i px} = e^{i(x \pm \sqrt{\Delta}) p} = \psi_{x\pm \sqrt{\Delta}}\;\;, \label{v-equation} 
\end{equation}

\noindent where $p$ is the conjugate momentum to the volume $V$.

Consequently, the holonomies build the braneworld quantum geometry as a regular lattice \cite{Silva:2020bnn}:

\begin{equation}
\gamma_{\sqrt{\Delta}} = \{x \in \mathbb{R}  \mid x = n\sqrt{\Delta}, \forall n \in \mathbb{Z} \}\;\;, \label{lattice}
\end{equation}

\noindent where \footnote{We have corrected by a factor $4$ the expression given to $\Delta$ in the reference \cite{Silva:2020bnn}.
Such correction comes from the fact that in the reference \cite{Singh:2015jus}, the universe critical density in the Friedmann equation that is consistent with our Hamiltonian appears multiplied by $4$. A similar correction has been done in the Eq. \eq{xi-sigma}.
} \cite{Singh:2015jus}

\begin{equation}
\Delta  = \frac{12\pi}{\sigma} \label{r-delta} \;,
\end{equation}


\noindent and $\gamma_{\sqrt{\Delta}}$ is a graph that corresponds to a $U(1)$ spin network, as those appear in the context of Loop Quantum Cosmology (LQC) \cite{Mielczarek:2012pf}. 

In this way, from the results obtained in \cite{Silva:2020bnn}, a Randall-Sundrum II brane describing a flat FRLW universe living on the $AdS$ bulk boundary must be seen as a $U(1)$ polymer structure, similar to LQC spin networks. 
In such a context, the discreteness in the position $x$ we have in the Eq. \eq{lattice}, with discreteness parameter $\sqrt{\Delta}$, implies that superselection rules for the brane gravitational sector will be imposed, in such a way the universe will evolve through discrete increments of the scale factor $a$ (or some object defined as a function of it, such as an area or volume).

The superselection rules also affect the bulk physics since, as one can remember, the brane
couples gravitationally to the bulk by emitting or absorbing closed strings, whose couplings, $g_{s}$, can be related to the brane tension as $g_{s} \sim 1/\sigma$ \cite{Becker:2007zj}.
One finds out, from the Eq. \eq{r-delta}, the following relation

\begin{equation}
g_{s}  \sim \Delta     . \label{delta-g}
\end{equation}



Such a result gives us the relationship between two parameters belonging to different theories connected through holography.
On the bulk side,  quantum information is encoded in the string coupling which tells us how closed strings interact.
On the boundary side, quantum information appears encoded in pixels with area $\Delta$. It is in agreement with the commensurability between qubits and quanta of area demonstrated in the context of the holographic principle \cite{Zizzi:2000jk}.


By considering a discrete spacetime evolution in the boundary theory, we shall have from the Eq. \eq{delta-g} that the string coupling will be constrained to have only nonvanishing finite values. In this way, the string modes which could lead to the dilaton divergency and, as a consequence, to the big bang singularity in the $AdS/CFT$ scenario \cite{Bak:2006nh, Engelhardt:2015gla}, are cut out. In this case, the initial singularity will be replaced by a bounce, as occurs in LQC \cite{Bojowald:2001xe, Ashtekar:2006rx, Bojowald:2006da}.


A second interesting point is that the polymer structures describing the boundary theory possess an advantage over the LQC ones by the fact they are defined by the brane tension, and not by the Barbero-Immirzi parameter. It turns the problem of Immirzi ambiguity \cite{Rovelli:1997na} absent in the scenario proposed in \cite{Silva:2020bnn} since the brane tension can be dynamically determined \cite{Becker:2007zj}. Such a result matches the idea proposed by several authors that a possible solution to the Immirzi ambiguity
can be obtained through a dynamical determination of the Barbero-Immirzi parameter \cite{Jacobson:2007uj, Taveras:2008yf, Mercuri:2009zi}.

\section{- A relational approach to quantum gravity: branes as quantum reference frames.}\label{sec3}

The scope of the results found out in \cite{Silva:2020bnn} may be even greater. It is because they can give us a new perspective about what may be a sui-generis aspect of quantum gravity: that spacetime must be not fundamental, as occurs in general relativity, but ought to emerge from quantum information. Such an idea has been worked out in the context of the $AdS/CFT$ correspondence \cite{Ryu:2006bv, Ryu:2006ef, Nishioka:2009un, Takayanagi:2012kg, VanRaamsdonk:2009ar, VanRaamsdonk:2010pw}, LQG \cite{Girelli:2005ii}, and other approaches \cite{Zizzi:2000jk}.

Particularly, in the LQG context, it has been argued that in a background independent quantum theory of gravity, spacetime must emerge as a collective phenomenon, from a deeper structure consisting only of quantum information encoded in the quantum correlations among quantum reference frames \cite{Aharonov:1984zz, Toller:1996ki, Rovelli:1990pi, Bartlett:2007zz, Giacomini:2017zju}.
In such a relational approach, if we have the right pattern of correlations, the system can be cleaved into parts that can be identified as different portions of spacetime. The degree of entanglement among such parts defines the notion of spatial distance:  the greater the entanglement between two portions of spacetime, the closer they are.

In such a context, the suitable definition of the quantum reference frames is fundamental to the construction of a quantum relational theory of spacetime. In this sense, from the results obtained in \cite{Silva:2020bnn}, branes can be conceived as such reference frames since they appear as $U(1)$ holonomy structures when one takes into account the perspective of the boundary theory in the $AdS/CFT$ correspondence. Specifically, due to their holonomy $U(1)$ framework, branes can be conceived as quantum clocks, i.e., temporal quantum reference frames \cite{Girelli:2005ii, Castro-Ruiz:2019nnl, Smith:2020zms}.




However, a detail we must add here is that holonomies can be used not only to establish branes as quantum reference frames but also to weave the entanglement network among them.
Actually, it has been demonstrated that holonomies consist of unitary transformations connecting two quantum geometrical states, belonging (in general) to two different Hilbert spaces, through entanglement \cite{Baytas:2018wjd, Bianchi:2018fmq, Mielczarek:2018jsh}:

\begin{equation}
\hat{h}_{ij} \; : \; \mathcal{H}_{i} \rightarrow \mathcal{H}_{j}\;,
\end{equation}

\noindent where

\begin{equation}
\hat{h}_{ij}\hat{h}_{ij}^{\dagger}  = \mathbb{I} \;.
\end{equation}

\vspace{5mm}


If we label the $\mathcal{H}_{i}$s as the Hilbert spaces of the branes in our collection, the indices $i,j$ above will range as $i,j = 1,...,N$,
and the holonomies $\hat{h}_{ij}$ must be generalized to $U(N)$ unitary matrices giving us the combinatorial rules related to the quantum geometric states describing the system of branes.

In such a relational scenario, $h_{ii}$ will give us simply the holonomies frozen to define the branes, while $h_{ij}$ ($i \neq j$) will describe the connections among them. In this way, one must observe that important detail of our construction is that there is not a fundamental distinction between the holonomies describing the quantum reference frames and the holonomies describing the quantum correlations connecting such frames. Consequently, even which we have called quantum references frames in the present paper can be understood, fundamentally, as quantum correlations. We shall return to such discussions in the conclusions section.

\section{- Branes as quantum reference frames and abstract spin networks.}\label{sec4}

 To trace out the combinatorial rules among the brane quantum clocks that build the boundary theory, we have an interesting fact pointed out by Girelli and Livine in \cite{Girelli:2005ii}. According to such a fact, a vector in the representation space, $\textit{\textbf{u}}(N)$, of the unitary group corresponds to an intertwiner with $N$ legs dressed by $SU(2)$ representations.




In this way, one can write the $U(N)$ group elements $\hat{h}_{ij}$ describing the system of $N$ branes in terms of $\textit{\textbf{u}}(N)$ basis vectors as

\begin{equation}
 \hat{h}_{ij} = e^{M_{ij}} = \mathbb{I} + M_{ij} + \frac{1}{2}M_{ij}M_{ik} +  ...\;\;\;,  \label{spin-net-exp}
\end{equation}

\noindent where $M_{ij} \in \textit{\textbf{u}}(N)$ corresponds to a $N$-leg intertwiner, in a way that the description of the boundary quantum geometry will be given as a superposition of intertwiner states (see Fig. \eq{fig1}). In such a context, we note that the number of intertwiners external legs will coincide with the number of branes sourcing the bulk geometry.

\begin{figure}[htb]
     \centering  
      \includegraphics[width=6.5cm, height=3cm]{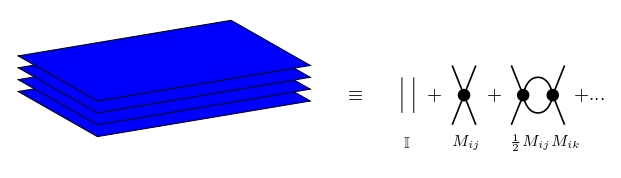}
      \caption{A collection of $N$ branes sourcing the bulk geometry is described from the boundary perspective as a superposition of $N$-leg intertwiner states. Each graph in the figure corresponds to a term belonging to the expansion on the right-hand side of the Eq. \eq{spin-net-exp} and describes a possible quantum state for the geometry of the $AdS$ boundary. } \label{fig1}
\end{figure}

By writing the $U(N)$ theory introduced in the last section in terms of an intertwiner basis, one can codify the quantum geometrical information related to the collection of $N$ branes into the $SU(2)$ representations dressing the intertwiner legs.
Since such quantum information appears encoded in pixels on the branes, one can attach the areas of such pixels to the intertwiner legs by expressing them in terms of the $SU(2)$ representations \footnote{The most simple relation between area and a $SU(2)$ representation has been traced in \cite{Krasnov:1998mp}, and used by Girelli and Livine \cite{Girelli:2005ii},
which fixes $A \sim j$.}.
In this case, intertwiners will give us the rules to weave the brane pixels according to $SU(2)$ selection rules. It corresponds to the basic idea of full LQG spin networks \cite{Girelli:2005ii}, which can be seen as an entanglement network relating quantum area states \cite{Shao:2010zza}.




However, we must observe that the intertwiners introduced in the present section correspond to abstract spin networks. Such objects, as highlighted by Girelli and Livine, are graphs labeled with $SU(2)$ representations, in the same way, we have in full LQG, but carrying only combinatorial information, without any reference to a background geometry or topology. 
Consequently, the theory introduced in the present section has been developed on a pure quantum informational perspective, consisting of a pre-spacetime, and consequently background-independent, approach.  


%

From the results we have found out,  the string/loop holographic relationship introduced in \cite{Silva:2020bnn} can be generalized as: a type IIB string theory living in a classical $AdS$ bulk emerges from a combinatorial $U(N)$ matrix theory described by full LQG (abstract) spin networks living on the bulk boundary, in the large $N$ limit. In such a context, the free parameters of the theories are related as

\begin{equation}
96\pi^{4}\alpha'^{2}g_{s}  = \frac{12\pi}{\sigma} = \Delta \;, \label{delta-g1}
\end{equation}

\noindent  and

\begin{equation}
\frac{L}{ l_{10}} = (4\pi N)^{1/4} \;. \label{curvature-spinnet}
\end{equation}

In the Eq.\eq{curvature-spinnet},

\begin{equation}
l_{(10)} = g_{s}^\frac{1}{4}\alpha'^{1/2} \; \label{10-planck}
\end{equation}

\noindent is the $10$-dimensional Planck length \cite{Becker:2007zj}. Moreover, $\alpha'$ is the Regge slope parameter.

Some comments are in order. At first, the Eq. \eq{delta-g1} has been borrowed from the results of the reference \cite{Silva:2020bnn}, where it appears from an extension of the so-called Black Hole Holographic Conjecture \cite{Emparan:2002px, Gregory:2004vt}. On the other hand, the Eq. \eq{curvature-spinnet} has been borrowed from the usual form of the $AdS/CFT$ correspondence. Such an equation gives us the $AdS$ radius $L$, which defines the bulk geometry \cite{Becker:2007zj}, in units of the ten-dimensional Planck length.
However,  Eq. \eq{curvature-spinnet}  assumes a new interpretation in the present context. It is because now $N$ turns to be given by the number of external legs in the intertwiners states whose determination depends on the combinatorial rules that govern the boundary theory. 

Consequently, quantities defined in terms of the $AdS$ radius turn to assume a combinatorial meaning. For example, using the Eqs. \eq{delta-g1}, \eq{curvature-spinnet} and \eq{10-planck}, we have that the $AdS$ scalar curvature \cite{Ammon:2015wua} can be expressed as

\begin{equation}
R = -\frac{12}{L^{2}} =  -24\pi\sqrt{\frac{6\pi}{N\Delta}}\;.
\end{equation}

\noindent In this way, curvature appears as a measure of how many quantum reference frames, carrying a quantum of area $\Delta$, are needed for the emergence of spacetime. Such a number will be determined by the combinatorial rules of the theory governing the pre-spacetime regime.

Second, we must note that the pre-spacetime regime described in the present section must occur far from the limit where the number of branes sourcing the bulk geometry is large, since in such a limit classical spacetime must emerge, according to the $AdS/CFT$ correspondence.
From the coincidence between the number of branes and the number of intertwiners external legs found out in the present section, we have that such reasoning is in agreement with that introduced in \cite{Girelli:2005ii} about the semiclassical limit of a pre-geometrical $U(N)$ matrix theory written in terms of $N$-leg intertwiner states.

\section{- The cosmological constant in the $AdS/CFT$ semiclassical scenario}\label{sec5}

The results obtained in the last section have linked the emergence of an $AdS$ bulk geometry to the combinatorial nature of the spin networks describing the $AdS$ boundary in a pre-spacetime regime. Such a relationship can be useful to discuss several issues in the context of quantum gravity. In the present section, we shall use it to address the cosmological constant problems.

In this sense, we have that, in the single brane case addressed in \cite{Silva:2020bnn}, the vacuum energy density is given in the semiclassical regime by \footnote{We have taken $\beta = l/l_{AdS} = \sqrt{4\pi\sigma/3}l$ in the Eq. (22) of the reference \cite{Silva:2020bnn}. Moreover, by fixing a typo in \cite{Silva:2020bnn}, we have taken $A_{l} = 4\pi\sqrt{2}l^{2}$, which gives us a additional factor $\pi^{-1}$ in the vacuum energy density $\lambda$.}:

\begin{eqnarray}
\lambda &=& 2\sigma\Big[\Big(1+\frac{9}{16\pi^2 l^{4}\sigma^{2}}\Big)^{1/2} -1  \Big] \nonumber \\
            &\approx& \frac{9}{16\pi^{2} l^{4} \sigma}\; \;,  \label{cosm-cont2}
\end{eqnarray}

\noindent where $l$ corresponds to a cutoff length scale.  

The result above has been obtained in the context of a flat FRLW universe living on the brane, for which the string/loop relationship has been proved to be valid. We note that a good amount of evidence for a flat universe has been pointed out, starting from the BOOMERanG experiment \cite{deBernardis:2000sbo}.

Now, by considering the usual $AdS/CFT$ scenario \cite{Maldacena:1997re} as the semiclassical limit of the theory developed in the last sections,  we shall have a collection of $N$ coincident identical branes, with $N$ large. In this situation, one must take $\sigma \rightarrow N\sigma$ in the Eq. \eq{cosm-cont2}, and obtain

\begin{eqnarray}
\lambda_{renorm} = \frac{9}{16\pi^{2} l^{4} N\sigma} = \frac{\lambda}{N} \label{cosm-cont3} \;.
\end{eqnarray}


%

From the equation above, the renormalized vacuum energy density will depend on the number of branes that source the bulk geometry. 
In this way, to determine the value of the cosmological constant, we now face a problem of braneworld engineering: how many branes do we need to build the universe?

The answer to such a question can be found out in the combinatorial rules governing spin networks, since we have learned from the results of the last section that the number of branes sourcing the bulk geometry is given by the number of external legs of the spin network that describes our universe, i.e., by the number of spin network legs piercing the universe horizon.

However, before one tries to get such a number, we need at first to define what we mean by "the universe horizon". In this way, we shall consider our universe as the object that carries the full holographic information necessary to the emergence of spacetime. In this sense, we have a hint from the $AdS/CFT$ correspondence, which tells us that causal diamonds defined on the $AdS$ boundary hold the necessary holographic information for the emergence of the bulk \cite{Hubeny:2012wa, Freivogel:2013zta, Headrick:2014cta}. Actually, due to their fixed lightcone structure, causal diamonds have played an important role in the literature on the holographic principle, being considered as the natural unit of holography in a region of spacetime \cite{Krishnan:2019ygy}.

Based on such considerations, we shall choose, as the universe horizon, the boundary of its causal diamond. 
Such a causal diamond consists of the entire region that can be probed by an observer inside the universe. 
It  is defined as the intersection of the past and future lightcones of the observer's worldline. In the present context, such a choice will trace out a scenario where the entire universe's causal structure emerges from the collection of brane quantum clocks.

The causal diamond of the universe has a fixed geometry, with a fixed radius given by

%
\begin{eqnarray}
R_{\Diamond} &=& \frac{1}{H_{0}}\int^{\infty}_{-1}\frac{dy}{\sqrt{\Omega_{m0}(1+y)^{3} + \Omega_{\Lambda 0}}} \nonumber \\
           &=& \Big(\frac{3}{8\pi\rho_{\Lambda 0}}\Big)^{1/2}\int^{\infty}_{-1}\frac{dy}{\sqrt{\frac{\Omega_{m0}}{ \Omega_{\Lambda 0}}(1+y)^{3}+1}}\;, \label{H-P1}
\end{eqnarray}
\noindent where $H_{0}$ is the current value of the Hubble parameter and $\rho_{\Lambda 0}$ is the current value of the vacuum energy density, which we will take as $\lambda_{renorm}$. 


In the Eq. \eq{H-P1}, $\Omega_{m0}$ and $\Omega_{\Lambda 0}$  correspond, respectively, to the density of matter and  density of cosmological constant parameters, defined as

\begin{equation}
\Omega_{m0} = \frac{8\pi}{3H_{0}^{2}}\rho_{m0}\;\;,\;\;  \Omega_{\Lambda 0} = \frac{8\pi}{3H_{0}^{2}}\rho_{ \Lambda 0}\;, \label{omega-def}
\end{equation}

\noindent where $\rho_{m0}$  is  the current value of the universe matter density.

Moreover, in the present dark energy dominated epoch, $\Omega_{m0}$  and $\Omega_{\Lambda 0}$ obey the following condition, in the case of a flat universe \cite{deBernardis:2000sbo}:

\begin{equation}
\Omega_{m0} + \Omega_{\Lambda 0} = 1\;. \label{balance}
\end{equation}


By fixing the essencial issue about the definition of the universe horizon, we are now in the position of putting the combinatorial rules governing spin networks to work. At first, we have that by considering the spin network legs piercing the universe horizon, the most important microstates consistent with a given area are those for which the $SU(2)$ representation carried by each spin network leg is as small as
possible. In this way, one would expect to consider spin $1/2$ legs. It agrees with the fact that, in the $AdS/CFT$ scenario, $N$ must be as large as possible to ensure the existence of a classical bulk spacetime.

Related to this, we shall take into account another combinatorial feature of spin networks, the so-called LQG projection constraint, which consists of the quantum analog of the
Gauss-Bonnet theorem, and arises from the consistency conditions for having a quantum horizon with the topology of a two-sphere \cite{Corichi:2006bs}. Such a constraint implies that a spin $1/2$ leg must pierce the
horizon at least twice, carrying in each piercing a half of a quantum of area on the universe boundary \cite{Pigozzo:2020zft, Carneiro:2020uww}.

Based on such considerations, we have that the total number of spin networks legs piercing the universe horizon can be calculated as

\begin{equation}
N  = 2\frac{A_{\Diamond}}{a_{min}} , \label{N-A}
\end{equation}

\noindent where $A_{\Diamond}$ corresponds to the area of the universe horizon, i.e., the area of the boundary of the universe causal diamond. Moreover,  $a_{min}$ is the quantum of area defined on it. 
The factor 2 appears as a consequence of the projection constraint.



In this way, from the Eqs. \eq{cosm-cont3} and \eq{N-A} we have,  by taking  $a_{min} = \Delta$, where $\Delta$ is given by the Eq. \eq{r-delta},

\begin{equation}
R_{\Diamond} = \Big(\frac{3}{2\rho_{\Lambda 0}} \Big)^{1/2}\;. \label{H-P2}
\end{equation}

\vspace{5mm}

\noindent To obtain the result above, we have also set $\lambda = \sigma$, i.e., we have interpreted the brane tension as the vacuum energy density of a single brane embedded in the $AdS$ bulk. It is just what is done in the usual Randall-Sundrum scenarios when one considers the point of view of an observer living on a brane \cite{Randall:1999ee, Randall:1999vf, Gron:2007}.


%
%
%

%
%

Now, by using the Eqs. \eq{H-P1} and \eq{H-P2}, we find out

\begin{equation}
\int_{-1}^{\infty}\frac{dy}{\sqrt{\frac{\Omega_{m0}}{ \Omega_{\Lambda 0}}(1+y)^{3}+1}} =  2\sqrt{\pi}\;. \label{H-L}
\end{equation}



%
%
%

%
%
%
%
%

By taking the Eqs. \eq{balance} and \eq{H-L}, the values of $\Omega_{m0}$ and $\Omega_{\Lambda 0}$ that correspond to the solutions of such a system of equations are:

\begin{equation}
\Omega_{m0}  = 0.3311 ,      \;\;\;\; \Omega_{\Lambda 0} = 0.6689 \;. \label{omegas-values} 
\end{equation}

Moreover, by using the Eq. \eq{omega-def} and $H_{0} = 67.36  \; km \; s^{-1}Mpc^{-1}$ \cite{Aghanim:2018eyx}, we obtain for the present value of the cosmological constant:

\begin{equation}
\Lambda_{0}  = 3H^{2}_{0}\Omega_{\Lambda_{0}} =  2.779  \times 10^{-122}\;, \label{lambda-value}      
\end{equation}

\noindent in Planck units. In the results above, the first three digits are effective.

Such results, obtained through the use of the idea that spacetime must emerge from quantum information, together with a fine combination of ingredients belonging to string theory and LQG, are in a good approximation with that given by the Planck 2018 cosmological data \cite{Aghanim:2018eyx}.

\section{- Conclusions and discussions}\label{conc}

\vspace{2mm}

We have generalized the results obtained in \cite{Silva:2020bnn} to undertake the idea that spacetime must emerge from quantum information. It led us to a relational approach where branes can be conceived as quantum clocks connected through quantum entanglement,
and the quantum description of the boundary theory in the $AdS/CFT$ correspondence turns to be given by a $U(N)$ matrix theory, which can be written in terms of full LQG (abstract) spin networks. 




Possible observational evidence for such a scenario has been presented, where  values for the matter and cosmological constant density parameters has been obtained, in a good approximation with the Planck 2018 results \cite{Aghanim:2018eyx}. It was possible by taking the usual $AdS/CFT$ scenario as the semiclassical limit of the theory. Such a result yields the balance between matter and dark energy we have in the present epoch. Moreover, by taking the current value of the Hubble parameter as given by the Planck 2018 collaboration, it has been possible to obtain an estimation for the cosmological constant in agreement with observations.



The results obtained in the present paper can be seen as the first phenomenological palpability for quantum gravity up to now relying on what may be a sui-generis feature of it: that spacetime must emerge from information encoded into quantum correlations. In this point, we note that,  since there is not a fundamental distinction between the holonomies describing the quantum reference frames and the holonomies describing the quantum correlations connecting such frames, even which we have called quantum references frames in the present paper can be understood, fundamentally, as quantum correlations.
In this way, only correlations matter in the present scenario, not correlations among things, but only correlations, in the Mermin sense \cite{Mermin:1998cg}. It differs from the usual $AdS/CFT$ perspective, where spacetime geometry may emerge from the correlations among entangled particles, and quantum correlations are distinct from the objects connected by them.

Moreover, our analysis differs from other approaches based on the holographic principle, as those appear in the so-called holographic cosmology \cite{Cohen:1998zx}, and in investigations related to de Sitter entropy \cite{Banks:2000fe}. Even though
in such approaches we have similar estimations to that given by the Eqs. \eq{cosm-cont3} and \eq{N-A}, one must note the peculiarity of the factor $2$ in the Eq. \eq{N-A}, coming from the LQG projection constraint, in our analysis. Such a factor has been provided by the combinatorial nature of the quantum gravity theory that underlies the count \eq{N-A}.

We note that such underlying quantum gravity theory is lakying in the aforementioned approaches \cite{Bousso:2000nf}, in a way that they may be blind for some details related to the microscopic description of spacetime that could be important for the calculation of the cosmological constant. In fact, such approaches are able only to propose an upper bound to the cosmological constant, which is not enough to answer the question of why it possesses a small but non vanish value. Moreover, it has been argued that the most probable value for the cosmological constant in the holographic cosmology scenario would be zero \cite{Horava:2000tb}, in evident disagreement with observations.

In this way, in addition to the introduction of a theoretical value for the cosmological constant that is in a good approximation with observations, the breakthrough of our work consists of providing a quantum gravity theory to support such a value, whose degrees of freedom are related to entanglement correlations, codified in the form of spin networks. 
Actually, such results show that the quantum gravity theory based on the string/loop holographic relationship firstly proposed in \cite{Silva:2020bnn}, and generalized in the present paper, has the potential not only to bring together the triumphs already obtained by string theory and LQG but also to solve problems that none of them have managed to solve so far, like the important issue of the cosmological constant.

Some points related to our results deserve further investigation. At first, to obtain the cosmological constant in the Eq. \eq{lambda-value}, we have used the current value of the Hubble parameter given by the Planck 2018 collaboration \cite{Aghanim:2018eyx}. The intermediate quantum gravity calculations, that lead to values of $\Omega_{m0}$ and $\Omega_{\Lambda 0}$ in good approximation with observations, do not require us to disturb the relationship between the cosmological constant and the Hubble parameter. It can be seen as an interesting aspect of the present paper. However, the dependence of $\Lambda_{0}$ on $H_{0}$
turns the question of why the cosmological constant has its value into the question of why the Hubble parameter has its current value. In this way, discussions about how fast is the universe expanding may be even more puzzling, especially in the face of the recent tension that has arisen in cosmology regarding the Hubble parameter \cite{Verde:2019ivm, DiValentino:2021izs}.


As a second point, we note that the discussion introduced in the present paper does not advocate a duality between two different theories anymore, as it has been done in \cite{Silva:2020bnn}, but the emergence of one theory from another, i.e., that string theory emerges from a pre-geometric theory described by spin networks.
In this case, some peculiar features belonging to superstring theory, like extra dimensions or supersymmetry, do not need to have a microscopic counterpart described by LQG, but they can be seen as emergent phenomena.  It may ripen the claims made in the reference \cite{Silva:2020bnn}.

In this sense, it is interesting to note that several papers in the literature have supported the idea that extra dimensions must emerge from a fundamental four-dimensional quantum gravity theory, starting from the seminal work by Arkani - Hamed, Cohen, and Georgi \cite{Arkani-Hamed:2001kyx}. Moreover, examples of emergent spacetime supersymmetry have been found out in the context of condensed matter physics \cite{Grover:2013rc}, where space-time supersymmetry naturally emerges at low energy and at long distances, although
the microscopic ingredients of the theory are not supersymmetric. Future investigations must shed more light on the role of supersymmetry and extra dimensions in the present context.

Another point  that deserves further discussion is that related to how classical spacetime can emerge from quantum correlations.
In this sense, we observe that  in the pre-spacetime regime, where one may have in general a collection of different quantum clocks, there must not be a defined metric, nor a defined causal structure, but they must depend on the quantum reference frame one takes \cite{Castro-Ruiz:2019nnl}. 
In this way, by considering the Copenhagen interpretation of Quantum Mechanics, one could evoke the necessity of an external super observer to our universe, and think about the emergence of spacetime as a result of a measurement performed by such an observer. Since each quantum clock is defined, as a quantum reference frame, by its internal time, in such a measurement process, all quantum clocks inside the universe would synchronize with the clock of the super observer, and consequently, synchronize among themselves. In such a scenario, a defined metric with a defined causal structure for the universe would be possible.


However, recent works have proved that the synchronization of quantum clocks can be implemented without the necessity of an external super observer, by using only the entanglement correlations among them \cite{Jozsa:2020}. In this context, 
the precision of the synchronization process depends on the entanglement amount we have \cite{Wang:2019aqr}.
In such a way, the emergence of classical spacetime may occur by the synchronization of the brane quantum clocks when the number of quantum correlations becomes large, with each correlation carrying the maximal amount of entanglement.

It fits the situation where we have a stack of $N$ coincident identical branes, in the limit of $N$ large, i.e., the usual $AdS/CFT$ scenario, where classical spacetime emerges in such a limit. Since values for $\Omega_{\Lambda_{0}}$, $\Omega_{m_{0}}$, and 
$\Lambda_{0}$ in agreement with observations are possible in such a scenario, as has been demonstrated in the present paper, such cosmological parameters may be coined when classical spacetime emerges, including its entire causal structure, as quantum entanglement provides brane clocks with equal readings to build the same story.

\bmhead{Acknowledgments}

The author acknowledges the anonymous referee for useful comments and suggestions.

\end{document}